\documentclass[10pt,conference]{IEEEtran}
\IEEEoverridecommandlockouts
\usepackage{cite}
\usepackage{amsmath,amssymb,amsfonts}
\usepackage{algorithmic}
\usepackage{graphicx}
\usepackage{textcomp}
\usepackage{xcolor}
\usepackage[hyphens]{url}
\usepackage{breakurl}
\usepackage[linesnumbered,ruled,vlined,noend]{algorithm2e}
\usepackage{setspace}
\usepackage{tabularx}
\usepackage{multirow}
\usepackage{tablefootnote}
\usepackage{threeparttable}
\usepackage{mdframed}
\usepackage{booktabs}
\usepackage{color}
\usepackage{fancyhdr}
\usepackage{tcolorbox}
\usepackage[colorlinks=false,hidelinks]{hyperref}
\usepackage{balance}
\usepackage{url}
\usepackage{tikz}
\usepackage{subfigure}
\usepackage{pifont}
\usepackage{fontawesome5}
\def\BibTeX{{\rm B\kern-.05em{\sc i\kern-.025em b}\kern-.08em
    T\kern-.1667em\lower.7ex\hbox{E}\kern-.125emX}}
\begin{document}

\title{Challenges of Using Pre-trained Models: the Practitioners' Perspective}

\author{\IEEEauthorblockN{Xin Tan}
\IEEEauthorblockA{\textit{School of Computer Science and Engineering} \\
\textit{Beihang University}\\
\textit{State Key Laboratory of Complex \& Critical Software Environment}\\
Beijing, China \\
xintan@buaa.edu.cn}
\and
\IEEEauthorblockN{Taichuan Li$^\ast$}
\IEEEauthorblockA{\textit{School of Computer Science and Engineering} \\
\textit{Beihang University}\\
Beijing, China \\
taichuanli@buaa.edu.cn}
\and
\IEEEauthorblockN{Ruohe Chen$^\ast$}
\IEEEauthorblockA{\textit{School of Economics and Management} \\
\textit{Beihang University}\\
Beijing, China\\
ruohechen@buaa.edu.cn}
\and
\IEEEauthorblockN{Fang Liu$^\dag$, Li Zhang}
\IEEEauthorblockA{\textit{School of Computer Science and Engineering} \\
\textit{Beihang University}\\
\textit{State Key Laboratory of Complex \& Critical Software Environment}\\
Beijing, China \\
fangliu@buaa.edu.cn, lily@buaa.edu.cn}
}
\maketitle

\def\thefootnote{$\ast$}\footnotetext{These authors contributed equally to this work.}\def\thefootnote{\arabic{footnote}}
\def\thefootnote{$\dag$}\footnotetext{Corresponding author.}\def\thefootnote{\arabic{footnote}}

\vspace{-0.5cm}

\begin{abstract}
The challenges associated with using pre-trained models (PTMs) have not been specifically investigated, which hampers their effective utilization. To address this knowledge gap, we collected and analyzed a dataset of 5,896 PTM-related questions on Stack Overflow. We first analyze the popularity and difficulty trends of PTM-related questions. We find that PTM-related questions are becoming more and more popular over time. However, it is noteworthy that PTM-related questions not only have a lower response rate but also exhibit a longer response time compared to many well-researched topics in software engineering. This observation emphasizes the significant difficulty and complexity associated with the practical application of PTMs. To delve into the specific challenges, we manually annotate 430 PTM-related questions, categorizing them into a hierarchical taxonomy of 42 codes (i.e., leaf nodes) and three categories. This taxonomy encompasses many PTM prominent challenges such as fine-tuning, output understanding, and prompt customization, which reflects the gaps between current techniques and practical needs. We discuss the implications of our study for PTM practitioners, vendors, and educators, and suggest possible directions and solutions for future research.
\end{abstract}

\section{Introduction}

Pre-training is a technique that leverages large amounts of unlabeled data to train a deep neural model, which can then be fine-tuned for specific downstream tasks with less labeled data~\cite{deguang2021review}. Pre-training has achieved remarkable success in various domains, especially in natural language processing (NLP)~\cite{min2021recent} and software engineering (SE)~\cite{karmakar2021pre}, where models such as BERT~\cite{devlin2018BERT}, GPT~\cite{radford2018improving}, and T5~\cite{raffel2020T5} have set new state-of-the-art results on many benchmarks. Especially in recent years, the birth of artificial intelligence (AI) tools such as ChatGPT\footnote{\url{https://openai.com/blog/chatgpt}}, GitHub Copilot\footnote{\url{https://github.com/features/copilot}}, \textit{etc}., pushed the PTMs to a climax.

PTMs have seen tremendous success recently due to their capability to simplify complex tasks and improve performance through pre-training and fine-tuning. Take GitHub Copilot, a most widely used AI Coding Assistant tool, as an example: GitHub Copilot is powered by Codex~\cite{chen2021codex}, a generative pre-trained AI model created by OpenAI. It has been trained on natural language text and source code from publicly available sources. Developers who use Copilot report that they completed the task 55\% faster than the developers who did not use GitHub Copilot~\cite{Eirini22Research}. Given the immense potential of PTMs, many researchers and AI practitioners are trying to apply PTMs to their own tasks. 

Although PTMs are popular and effective, fully harnessing their potential can be challenging for practitioners. On the one hand, PTMs often have complex architectures and mechanisms, making it difficult for practitioners to gain a deep understanding of their inner workings~\cite{han2021pre}. On the other hand, predefined architectures and computational resource requirements can hinder practitioners from tailoring PTMs to their specific needs or incorporating domain-specific knowledge effectively~\cite{zhou2023comprehensive}. Worse still, PTM's community support is currently lacking in maturity because of its rapid development.  
These challenges are evident in the frequent questions raised by developers in QA forums regarding the use of PTMs. However, there has yet to be a systematic examination of these challenges, which hinders the effective utilization and improvement of PTMs.

To better understand practitioners' challenges and needs, we conduct a large-scale analysis of Stack Overflow (SO) questions related to pre-training. SO\footnote{\url{https://stackoverflow.com/}} is the most popular online platform where programmers can ask and answer questions about various topics in software development. It is a valuable source of data for researchers who want to study the trends, patterns, and challenges in SE~\cite{treude2016augmenting,haque2020challenges,chen2020comprehensive,wang2023automl}. Specifically, we collect 5,896 PTM-related questions from SO, covering a time span from 2018 to 2023. We perform various statistical and qualitative analyses on this dataset, aiming to answer the following research questions:

\begin{itemize}
    \item[RQ1] \textbf{Popularity Trend.} What is the popularity trend of PTM-related questions?
    \item[RQ2] \textbf{Difficulty.} What is the difference in difficulty between PTM-related questions and other questions? 
    \item[RQ3] \textbf{Taxonomy of Challenges.} 
    What specific challenges do PTM practitioners face? Are these challenges different from traditional deep learning challenges? 
\end{itemize}

To address RQ1, we extract and tally the occurrences of various PTM-related tags in the questions, ranking them according to their frequency of mention. To answer RQ2, we compare the response rate, response time, and ratio of answers to views of PTM-related questions and questions unrelated to PTMs, allowing us to assess the level of difficulty. Our findings reveal a consistent upward trend in the number of PTM-related questions over time. PTM-related questions also generally require significantly more time to answer, indicating a higher level of difficulty. To delve into the specific challenges (RQ3), we perform a thematic analysis~\cite{cruzes2011recommended} of the PTM-related questions, which allows us to extract and categorize the main challenges. We establish a comprehensive taxonomy consisting of three categories and 42 codes, linking challenges with Model Lifecycle Management, Coding-related Issues, and Others. Compared with the challenges faced by deep learning (DL) practitioners, PTM practitioners also face various unique prominent challenges, e.g., fine-tuning, output understanding, and memory management. 
The contributions are as follows:

\begin{itemize}

\item We present the first systematic analysis of PTM-related questions on SO, providing a comprehensive overview of the common challenges and needs of PTM practitioners. 

\item We provide valuable insights and guidance for PTM practitioners, vendors, and educators, and suggest future research directions and opportunities. 

\item We create and release a dataset of 5,896 questions related to PTMs, which is annotated with categories and metadata. The dataset can be used for further research. 
\end{itemize}


\section{Background}

\vspace{-0.2cm}
\subsection{PTMs}

PTMs are deep neural network models that are trained on large amounts of unlabeled data, such as text, images, or speech, to learn general and reusable representations or knowledge~\cite{wang2022pre}. These representations or knowledge can then be transferred to specific downstream tasks with less labeled data, by fine-tuning the PTMs with task-specific layers or objectives~\cite{han2021pre}. PTMs have witnessed remarkable advancements across different fields, particularly in NLP and SE domains~\cite{min2021recent}. Models like BERT~\cite{devlin2018BERT}, GPT~\cite{DBLP:journals/corr/abs-2005-14165}, and T5~\cite{raffel2020T5} have demonstrated exceptional performance, establishing new state-of-the-art benchmarks in these areas.

PTMs are categorized into two types based on pre-training objectives: generative PTMs and discriminative PTMs~\cite{han2021pre}. Generative PTMs learn the joint probability distribution of input data, such as words or pixels, and generate new data samples that follow the same distribution~\cite{oussidi2018deep}. They typically use autoregressive or autoencoding objectives to maximize the likelihood of input data~\cite{liu2021self}. For instance, GPT is a generative PTM used for text data, employing a transformer decoder to predict the next word given the previous words~\cite{radford2018improving}. Discriminative PTMs aim to learn the conditional probability distribution of output labels given the input data, such as sentiment polarity or object category~\cite{yang2021survey}. They usually adopt contrastive or cloze objectives to maximize the likelihood of the output labels~\cite{liu2021self}. For example, BERT is a discriminative PTM used for text data, utilizing a transformer encoder to predict masked words given the surrounding words~\cite{devlin2018BERT}.

\subsection{Pre-training Toolkits}
Pre-training toolkits are essential for PTMs by providing the necessary infrastructure and resources. They offer a comprehensive range of functionalities and interfaces that enable easy implementation and utilization of PTMs. Pre-training toolkits usually support various pre-training objectives, architectures, datasets, and downstream tasks, allowing users to customize their configurations and parameters. Pre-training toolkits also provide optimization techniques and distributed computing mechanisms to improve the efficiency and scalability of PTMs. Some popular pre-training toolkits include HuggingFace Transformers~\cite{wolf2019huggingface}, Fairseq~\cite{ott2019fairseq}, \textit{etc}.

\section{Data Preparation}\label{sec: data_preparation}
We focus on SO, which is the most trusted and popular QA website for programmers and has been an important study subject for many SE-related research~\cite{treude2016augmenting,zhang2019empirical,chen2020comprehensive}. As of March 2022, SO has over 20 million registered users and has received over 24 million questions and 35 million answers~\cite{wiki:Stack_Overflow}. Given its popularity among the developer community, SO is well-suited to analyze the challenges that practitioners face when applying PTMs. In this section, we describe how we collect related questions from SO. To collect the data, we used SO API\footnote{\url{https://api.stackexchange.com/}}. Considering the popular time span of PTMs, we established a data collection window spanning from January 1, 2018, to June 15, 2023. 

We utilized the snowball sampling technique~\cite{biernacki1981snowball} to gather PTM-related questions, initiating with an initial tag list and expanding it based on relevance. After brainstorming with all authors, who have over three years of experience with PTMs, we compiled a list of initial PTM-related tags. The initial set consisted of 10 tags and most of them are directly the name of popular PTMs (i.e., \textit{``bert-language-model'', ``gpt-2'', ``gpt-3'', ``roberta'', ``roberta-language-model'', ``bart'', ``t5-transformer'', ``huggingface-transformers'', ``huggingface-tokenizers'', ``huggingface-datasets''}). Through this approach, we collected a total of 4,303 PTM-related questions.
Then, we considered the related tags recommended by the SO platform itself to expand the initial tag set. We discussed each recommended tag together to decide whether it was a PTM-related tag. If a recommended tag was related to PTMs, we added it to our tag set and used it to collect more questions. We iterated this process until there were no PTM-related tags recommended by SO. Eventually, we collected 18 tags, mainly involving three types, i.e., names of certain PTMs, PTM provider -- \textit{Hugging Face}, and PTM vendor -- \textit{OpenAI}. Table \ref{tab:tags} shows the name and frequency of each tag. In total, we obtained 5,896 PTM-related questions. The manual annotation conducted for RQ3 (refer to Section \ref{sec:RQ3_approach}) demonstrates a low false positive rate (5.8\%), providing evidence for the validity of our tag selection process.

\begin{table}[h]
\centering
\scriptsize
\vspace{-0.3cm}
\caption{Names and Frequencies of PTM-related Tags}
\vspace{-0.1cm}
\begin{tabular}{|@{\hspace{1.5pt}}l@{\hspace{1.5pt}}|@{\hspace{1.5pt}}l@{\hspace{1.5pt}}|@{\hspace{1.5pt}}l@{\hspace{1.5pt}}|@{\hspace{1.5pt}}l@{\hspace{1.5pt}}|}
\hline
\bf{Selected Tags} & \bf{No.(Freq.)} & \bf{Selected Tags} & \bf{No.(Freq.)}\\
\hline
huggingface-transformers & 2,563 (43\%) & gpt-2 & 174 (3\%)\\
bert-language-model & 1,763 (30\%) & sentence-transformers & 142 (2\%)\\
openai-api & 1,053 (18\%) & fine-tune & 114 (2\%)\\
pre-trained-model & 423 (7\%) & openai-whisper & 107 (2\%)\\
huggingface-tokenizers & 413 (7\%) & roberta-language-model & 62 (1\%)\\
huggingface & 403 (7\%) & roberta & 32 (1\%)\\
gpt-3 & 245 (4\%) & gpt-4 & 22 (0.3\%)\\
chatgpt-api & 219 (4\%) & bart & 14 (0.2\%)\\
huggingface-datasets & 187 (3\%) & t5-transformer & 6 (0.1\%)\\
\hline
\end{tabular}
\label{tab:tags}
\end{table}

\vspace{-0.3cm}
\section{RQ1: Popularity Trend}

\subsection{Motivation} 
Despite the growing popularity of PTMs, we still do not know the fluctuating trends and interests surrounding this field on SO. This lack of understanding hinders our ability to gain valuable insights into the evolution of PTMs.

\vspace{-0.3cm}
\subsection{Approach} 
To analyze the popularity trend of PTMs, we utilized the PTM-related questions collected in Section~\ref{sec: data_preparation}. These questions were grouped by year and by tag, enabling us to determine the number of questions for each group. By doing so, we gathered the frequency of each tag within each year. This approach allowed us to gain insights into the evolving interests and trends surrounding PTMs over time.

\subsection{Results} 
\begin{figure}[h]
\centering 
\vspace{-0.3cm}
\includegraphics[scale=0.42]{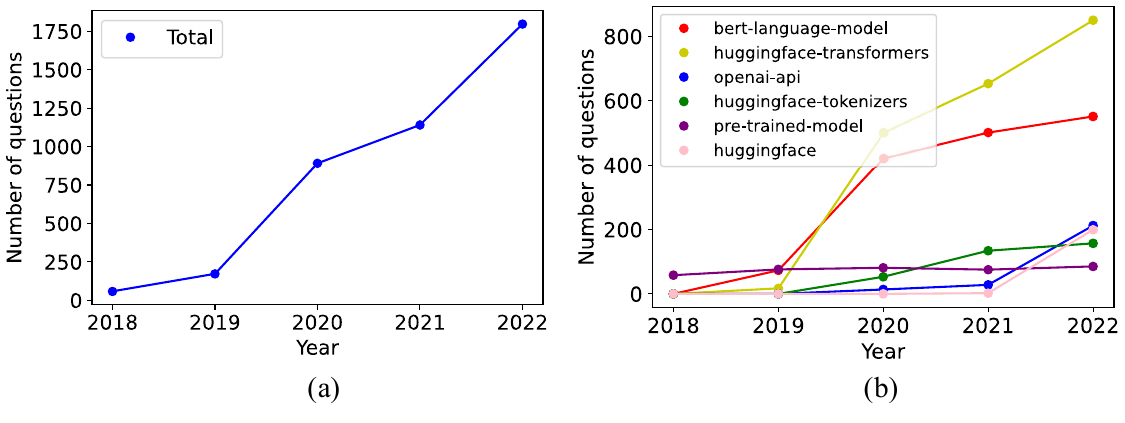} 
\vspace{-0.3cm}
\caption{Popularity Trend of PTM-Related Questions and Top Six Tags} 
\vspace{-0.3cm}
\label{fig:popularity trend} 
\end{figure}

Fig. \ref{fig:popularity trend} (a) depicts the popularity trend of PTM-related questions. The graph highlights a consistent upward trend, indicating the increasing popularity of PTMs over the years. Before 2019, the number of PTM-related questions was relatively low, reflecting a limited awareness and adoption of PTMs. However, as time passed by, there is a significant surge in the number of questions, indicating a growing interest and engagement with PTM technology. Notably, in 2022, there are more than 2,000 questions reported, reflecting the heightened attention and demand for information about PTMs. This substantial increase further reinforces the growing recognition and significance of PTMs in various domains. Fig. \ref{fig:popularity trend} (b) depicts the evolution of the six most popular tags, demonstrating a general upward trend. Specifically, the tags ``\textit{huggingface-transformers}'' and ``\textit{bert-language-model}'' experienced significant growth from 2019 to 2020. Similarly, the tags ``\textit{pre-tranined-model}'' and ``\textit{huggingface}'' saw substantial increases from 2021 to 2022.

\begin{mdframed}[linecolor=gray,roundcorner=12pt,backgroundcolor=gray!15,linewidth=3pt,innerleftmargin=2pt, leftmargin=0cm,rightmargin=0cm,topline=false,bottomline=false,rightline = false]
  \textbf{Summary for RQ1:} The number of PTM-related questions has been consistently increasing over the years, indicating the growing recognition and widespread adoption of PTMs, as well as the significant challenges that practitioners face when working with these models.
\end{mdframed}

\section{RQ2: Difficulty}

\subsection{Motivation}
RQ2 examines the difficulty of PTM-related questions in comparison to other topics. It can provide insights into the level of challenge, helping assess the need for additional support to enhance user experience.

\subsection{Approach}
In order to comprehend and quantify the difficulties encountered by developers when using PTMs, we employed the following three widely adopted metrics~\cite{ahmed2018concurrency, rosen2016mobile, yang2016security}:
\begin{itemize}
    \item Percentage of the questions with no accepted answer (i.e., \emph{\%no acc}).
    \item Response time required to receive an acceptable answer.
    \item The ratio of the average number of answers these questions receive to the average number of views these questions receive (i.e., \emph{PD})~\cite{yang2016security}.
\end{itemize}

To establish a baseline for comparison, we utilized questions posted on SO after January 1, 2018, excluding PTM-related questions, which we denoted as ``PTM-unrelated questions''. For the first metric, we conducted a proportion test~\cite{storer1990exact} to compare the percentages of questions with no accepted answers between PTM-related and PTM-unrelated questions. This statistical test is appropriate for comparing the \emph{\%no acc} values and evaluates the null hypothesis that the proportions are equivalent across multiple groups. Regarding the second metric, we analyzed questions with accepted answers and visualized the distribution and median response time for both PTM-related and PTM-unrelated questions. The third metric involves calculating the average values of ``AnswerCount'' and ``ViewCount'' for PTM-related and PTM-unrelated questions to obtain \emph{PD} values. Generally, when a question obtains a high number of views but receives only a limited number of answers, it suggests that only a small fraction of the audience possesses the expertise to respond adequately~\cite{yang2016security}. Thus, a lower score of \emph{PD} indicates a higher level of difficulty for questions within that topic.

\subsection{Results}
For the percentage of questions without an accepted answer (\emph{\%no acc}), we find that PTM-related questions have a value of 73.9\%, while PTM-unrelated questions have a value of 57.8\%. The proportion test results ($\chi^2= 628.830, df = 1, \text{p-value} < 1e-100$) confirm the significant difference, indicating that PTM-related questions are more challenging to answer compared to PTM-unrelated questions. Table \ref{tab:no_acc and PD} shows \emph{\%no acc} of the top six popular tags. 
We observe that the questions tagged with ``\textit{openai-api}'' and ``\textit{huggingface}'' exhibit the highest values of \emph{\%no acc}, indicating that they are most difficult to answer.
In terms of this metric, PTM-related questions also present greater difficulties than well-researched topics in SE, such as big data (\emph{\%no acc} = 60.5\%)~\cite{bagherzadeh2019going}, concurrency (\emph{\%no acc} = 43.8\%)~\cite{ahmed2018concurrency}, and mobile devices (\emph{\%no acc} = 55.0\%)~\cite{rosen2016mobile}.

\begin{table}[h]
\centering
\scriptsize
\vspace{-0.3cm}
\caption{\emph{\%no acc} and \emph{PD} of PTM-related Tags}
\vspace{-0.2cm}
\begin{tabular}{|l|l|l|}
\hline
\bf{Selected Tags} & \bf{\emph{\%no acc}} & \bf{\emph{PD}} \\
\hline
huggingface-transformers & 0.70 & 0.05 \\
bert-language-model & 0.74 & 0.05 \\
openai-api & 0.78 & 0.08 \\
pre-trained-model & 0.68 & 0.06 \\
huggingface-tokenizers & 0.70 & 0.05 \\
huggingface & 0.78 & 0.12 \\
\hline
\end{tabular}
\label{tab:no_acc and PD}
\end{table}

\begin{figure}[h] 
\centering 
\vspace{-0.3cm}
\includegraphics[scale=0.4]{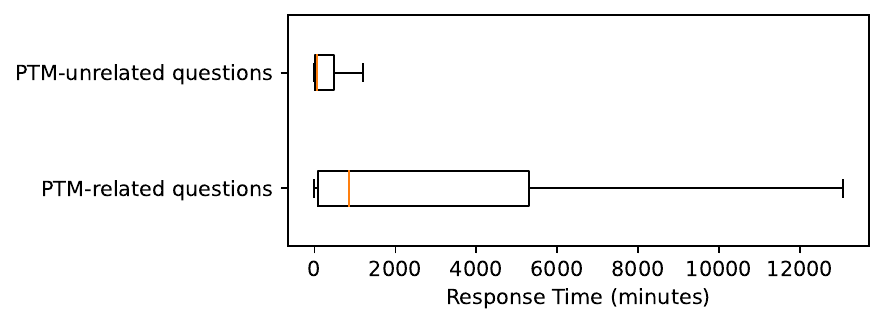} 
\vspace{-0.3cm}
\caption{Time Needed to Receive an Accepted Answer} \label{fig:difficulty} 
\vspace{-0.3cm}
\end{figure}

Fig. \ref{fig:difficulty} illustrates the distribution of response time required to receive accepted answers for PTM-related questions and PTM-unrelated questions. The response time for PTM-unrelated questions predominantly falls below 1,500 mins, while the distribution range is significantly wider for PTM-related questions. We also discover that the average response time for PTM-related questions is more than twice that for PTM-unrelated questions, with 22,676 mins and 9,745 mins, respectively. The median response time for PTM-related questions is 851 mins, which is over 15 times longer than 55 mins for PTM-unrelated questions. In terms of this metric, PTM-related questions also present greater difficulties than well-researched topics in SE, such as big data (198 mins)~\cite{bagherzadeh2019going}, concurrency (42 mins)~\cite{ahmed2018concurrency}, and mobile devices (55 mins)~\cite{rosen2016mobile}.

For the third metric, we obtained the \emph{PD} scores of PTM-related and PTM-unrelated questions as 0.06 and 0.12 respectively. Table \ref{tab:no_acc and PD} shows \emph{PD} of the questions with the top six popular tags. Even when compared to the most challenging subtopics in the field of Security (e.g., ``\textit{Java Security}'', ``\textit{Asymmetric Encryption}'', and ``\textit{Bug}''), where the minimum PD value is 0.07~\cite{yang2016security}, the difficulty level of PTM-related questions with certain tags remains lower. This suggests that answering PTM-related questions poses a greater challenge, requiring a deeper understanding and expertise.

\begin{mdframed}[linecolor=gray,roundcorner=12pt,backgroundcolor=gray!15,linewidth=3pt,innerleftmargin=2pt, leftmargin=0cm,rightmargin=0cm,topline=false,bottomline=false,rightline = false]
  \textbf{Summary for RQ2:} Our findings suggest that developers encounter significant challenges when utilizing PTMs, as indicated by the response rate, response time, and the ratio of answers to views. These challenges highlight the need for further research to better understand the specific difficulties underlying PTM-related questions. Such research can provide valuable insights and pave the way for addressing these challenges effectively.
\end{mdframed}

\section{RQ3: Taxonomy of Challenges}
\vspace{-0.2cm}
\subsection{Motivation}
As PTM-related questions gain increasing attention and prove to be difficult to resolve, it is crucial to investigate the particular challenges encountered by PTM practitioners, especially the unique challenges compared to general DL models. This information can reveal the distinct support required.

\vspace{-0.2cm}
\subsection{Approach}\label{sec:RQ3_approach}
\subsubsection{Collecting Questions for Manual Analysis} 
While we identified 5,896 PTM-related questions in Section~\ref{sec: data_preparation}, their quality cannot be guaranteed, directly impacting the quality of the constructed taxonomy. To address this, we refined our dataset by excluding PTM-related questions with a score below 1, determined by SO user votes. A question's score is calculated by subtracting the number of downvotes from the number of upvotes, with a higher score indicating better question quality~\cite{baltadzhieva2015predicting}. Through this process, we filtered out vague, unclear, trivial, or duplicated questions that were unsuitable for our study, resulting in 2,829 high-quality PTM-related questions. Manually analyzing 2,829 questions is time-consuming and can lead to researcher fatigue, risking decreased accuracy and attention to detail in the annotation process. To address this, we adopted two sampling strategies: random sampling and high-score sampling. \textbf{Random sampling} aimed to ensure a diverse and representative distribution of topics. \textbf{High-score sampling} aimed to complement the questions with high quality and great significance. We combined these two sampling strategies to balance the trade-off between diversity and significance.

Following previous work~\cite{chen2020comprehensive, zhang2019empirical}, we conducted a random sampling of 339 questions from 2,829 high-quality PTM-related questions, aiming to achieve a 95\% confidence level and a 5\% confidence interval~\cite{ci1987confidence}. This ensures that our sample is representative and reliable to reflect the characteristics of the PTM-related questions. Considering that random sampling may not cover the most important and high-quality questions, we decided to supplement the questions with a score greater than or equal to nine. This selection results in 116 questions accounting for 4.1\% of high-quality PTM-related questions. We opted for a score of nine as the threshold, strategically aiming to concentrate on the most exemplary issues while considering the feasibility of manual analysis. It is worth noting that there were 25 overlapping questions between these two sampling strategies. In total, we obtained 430 questions as our final sample for annotation and analysis. The size of this dataset is comparable to those used in previous studies~\cite{chen2020comprehensive, wang2023automl} that also required manual analysis of SO posts.

\subsubsection{Conducting Manual Analysis} To analyze the challenges within the questions, we conducted a thematic analysis~\cite{cruzes2011recommended}. 
We initially analyzed a subset of 129 questions (30\% of the total 430 questions) by carefully reading and re-reading them, including the title, body, code snippets, comments, and answers. We assigned initial codes to reflect the underlying challenges of these questions, using keywords or conducting further investigation for deeper understanding. We also reviewed answers and used the Google search engine to gather additional information. If a question is not related to PTM, we classify it as a False positive.

We then grouped similar codes into categories and created a hierarchical taxonomy of challenges. This iterative process involved identifying recurring patterns, concepts, and ideas within the questions and organizing them accordingly. The outcome was a comprehensive codebook illustrating various codes related to PTM challenges and their meaning. This analysis was performed by the first four authors together, i.e., two professors, and two undergraduate students, all with over three years of PTM-related experience. This collaborative approach ensured consistency and reliability in the pilot analysis, providing a data guarantee for the subsequent analysis.

Using the developed codebook, the remaining 70\% of questions were labeled independently by the second and third authors. In cases where a question did not fit within the existing taxonomy, the first four authors held discussions to establish new codes and enhance the codebook and taxonomy. In the later stage of the coding process, there were no new codes emerged, which means that we had reached data saturation~\cite{fusch2015we}. 
To assess the reliability of the labeling process, inter-rater agreement was measured during the independent labeling phase. The agreement, measured using Cohen's Kappa, was found to be 83\%, indicating perfect agreement and highlighting the reliability of our coding schema and procedure. Any conflicts or discrepancies in labeling were resolved through discussions and consensus among the first four authors, ensuring consistency and accuracy in the final classification of the questions.

\subsubsection{Mapping to DL challenges}  To comprehend the differences in challenges between PTM and general DL, we aligned our taxonomy with the work of Zhao et al.~\cite{zhao2021state}. They investigated the challenges of DL by analyzing SO and obtained 30 challenges. We identified the analogous challenges and annotated them accordingly. There are certain codes that do not align with our taxonomy. These codes primarily pertain to specific applications (e.g., ``\textit{Object Detection}'') or specific DL models (e.g., ``\textit{CNN Structure}'').

\subsection{Results}

\begin{figure*}[h]
  \centering
  \includegraphics[scale=0.43]{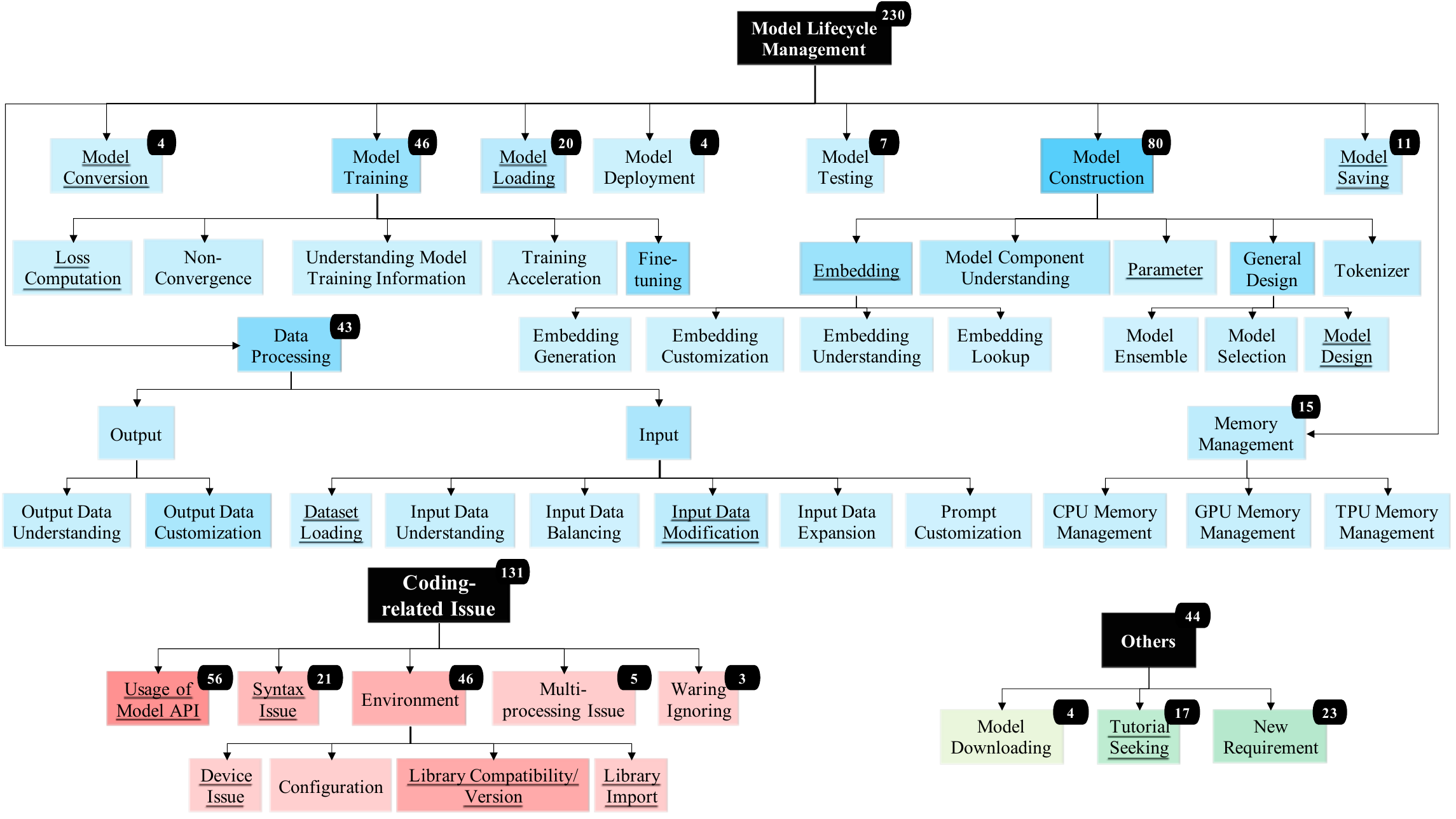}
  \vspace{-0.3cm}
 \caption{Taxonomy of PTM Challenges. The challenges highlighted in \underline{codes} indicate that these challenges are also prevalent in the broader field of deep learning~\cite{zhao2021state}, while others are prevalent in PTMs.}
 \vspace{-0.3cm}
  \label{fig: taxonomy}
\end{figure*}

Fig.~\ref{fig: taxonomy} shows the hierarchical taxonomy of challenges extracted from PTM-related questions. The taxonomy is structured into multiple layers with varying numbers across different categories. We identified a total of 42 challenges as leaf nodes within the taxonomy, which is composed of three primary categories. 
We can see that PTM practitioners also encounter common challenges in DL~\cite{zhao2021state}, such as \textit{Model API Usage}, \textit{Input Data Processing}, \textit{Embedding}, and \textit{Model Design}. However, due to the uniqueness of PTMs, PTM practitioners face many specific and prominent challenges, such as \textit{Fine-tuning}, \textit{Output Understanding}, and \textit{Prompt Customization}. We provide further details on these categories by combining the features of PTMs and giving examples.

\vspace{1mm}
\noindent \textbf{\textit{I. Model Lifecycle Management (56.79\%)}}
\vspace{1mm}

This is the most extensive and significant category, covering more than 50\% of the PTM-related questions. It consists of nine sub-categories related to the complete lifecycle of a PTM, from its construction to testing and deployment. 

\textit{\textbf{1) Model Construction (19.75\%).}} Constructing PTMs can pose challenges for many practitioners in the field. Notably, nearly a quarter of questions pertain to this subcategory. It encompasses various crucial aspects that directly impact the construction of PTMs, e.g., General Design, Model Component Understanding, and Tokenizer. Each of these topics plays a vital role in shaping the overall construction and functionality of PTMs.

\textbf{General Design}. Designing the appropriate architecture and approach for PTMs can be challenging for developers. One challenge is \textit{Model Design}, which involves making decisions on the neural network architecture, pre-training objectives, regularization techniques, and model size to effectively apply to downstream tasks. For example, a developer wanted to disable or freeze certain layers by modifying the model's architecture~\cite{Merle2023disable}. Another challenge is \textit{Model Selection}, as the wide range of available PTMs requires careful consideration of factors such as performance, domain similarity, computational requirements, available resources, etc. Developers usually have difficulty in selecting suitable models for their tasks, e.g., ``\textit{I want to use Spacy's pre-trained BERT model for text classification but I'm a little confused about cased/uncased models}''~\cite{Oleg2020Cased}.
Developers also feel difficulty in \textit{Model Ensemble} due to the need to consider compatibility, selection, and strategy for optimal integration of multiple PTMs. For example, ``\textit{I'm trying to create an ensemble with three pre-trained VGG16, InceptionV3, and EfficientNetB0 for a medical image classification task...But when I execute the code, I get this error...}''~\cite{Lamyaa2020Build}.

\textbf{Model Component Understanding}. Understanding the components of PTMs can be challenging, especially for developers new to working with such models. Developers often have issues related to the model's internal information, like concepts and properties. For instance, ``\textit{I've googled but didn't find a single guide that allowed me to view how a pre-trained torch neural network is designed/coded}''~\cite{Shenath2018Is}. Worse still, the internal workings of PTM are like a black box to practitioners, making it difficult to directly observe or explain the specific workings of the model. This makes it challenging to understand the functionality and interactions of these components.


\textbf{Tokenizer}. Developers face challenges related to tokenization when breaking down a sequence of text into smaller units called tokens. These challenges arise when applying tokenization to a specific task or when encountering errors during the tokenization process. For example, a developer asked ``\textit{how do I use ByteLevelBPETokenizer with UTF-8?}''~\cite{kloop2023How}. An example of a common error encountered during tokenization of BERT or DistilBERT models is the \textit{ValueError: "TextEncodeInput must be Union[TextInputSequence, Tuple[InputSequence, InputSequence]]}~\cite{Raoof2020ValueError}. his error suggests a problem with the input data format/type provided to the tokenizer. This question has gained significant attention, with 41k views and 35 upvotes. The top-rated answer, receiving 79 upvotes, suggests a solution of \textit{removing all None rows from DataFrame columns before converting them to a list}.

\textbf{Embedding}. An embedding is a numerical vector representation of words, sentences, or documents that captures the semantic and contextual information of the input text~\cite{li2018word}. Developers often encounter challenges related to embeddings, including \textit{Embedding Generation}, \textit{Embedding Understanding}, \textit{Embedding Customization}, and \textit{Embedding Lookup}.
For instance, a developer had a specific question about generating embeddings using BERT and asked, ``\textit{How to cluster similar sentences using BERT}''~\cite{somethingstrang2019How}. This question has received 34k views and 30 upvotes, indicating a need for examples and guidance on generating embeddings. Developers also face difficulties when customizing their embeddings. For instance, a developer mentioned, ``\textit{Earlier I've used Glove embedding to build the seq2seq model for text summarization. Now I want to change the Glove with BERT to see the performance of the model}''~\cite{Ganesh2020Using}. These examples highlight the importance of providing guidance for developers in generating embeddings and customizing them to suit their specific tasks.

\textbf{Parameter}. Only a few developers face challenges when working with PTM parameters. Developers often need to fine-tune or adapt the PTM parameters to their specific task or dataset. This involves adjusting hyper-parameters, e.g., learning rate, batch size, regularization method, and optimizer parameters like momentum, and learning rate decay. For instance, because some optimizer parameters are not specified correctly, the switch from a custom-built model to a pre-trained BERT model resulted in a significant drop in accuracy~\cite{Spacy2020May}.

\textit{\textbf{2) Data Processing (10.62\%).}} This subcategory is related to input and output data processing, accounting for over 10\% of PTM-related questions, indicating its significance in applying PTMs. This includes challenges such as dataset loading, input data modification, and handling the model's output to suit the specific needs of the application or downstream tasks.

\textbf{Input}. Developers encounter several challenges when processing input data for PTMs, with \textit{Input Data Modification} being a common obstacle. This involves tasks such as cleaning, preprocessing, formatting, and transforming the data to meet PTMs' requirements, which can be puzzling for developers. For instance, when using BERT to analyze tweet sentiment, a developer faced the challenge of identifying which elements to remove, e.g., \textit{@names, URLs, or numbers}~\cite{Skalonga2020Data}. Developers also struggle with \textit{Understanding Input Data}, including settings or shapes. For example, a developer questioned the need for a throw-away column in the BERT format~\cite{anegru2019Why}, highlighting the difficulty in comprehending certain data-related elements or configurations. \textit{Input Data Expansion} presents challenges as well. For example, due to time constraints, a developer sought to incorporate new data in order to continuously train the model instead of starting the training process from scratch~\cite{Tia2018oct}. Additionally, a few developers encounter challenges with \textit{Prompt Customization}, \textit{Dataset Loading}, and \textit{Input Data Balancing}. These challenges emphasize the complexity of working with the input data of PTMs and the diverse aspects that developers need to address. 

\textbf{Output}. Developers encounter challenges related to the output of PTMs. These challenges include \textit{Output Data Understanding} and \textit{Output Data Customization}. Output Data Understanding refers to the difficulty in comprehending the format, structure, or interpretation of the model's output. For instance, a developer asked about the meaning of the second output of Huggingface's BERT~\cite{user21828572020What}. Compared to DL, PTMs often generate complex output representations that require further analysis or processing to extract meaningful insights. Understanding the output data is crucial for effectively interpreting and utilizing the model's predictions. Output Data Customization involves the need to customize or adapt the model's output to fulfill specific requirements or downstream tasks. This process may involve post-processing, format conversion, or integrating additional information. For example, a developer encountered challenges in retrieving all documents related to a topic when using BERTopic for topic modeling, as it only returned three documents per topic~\cite{Kaleem2021How}.

\textit{\textbf{3) Model Training (10.36\%).}} PTM is usually pre-trained on large-scale corpora, but when applied to specific tasks, it needs to be adapted to real-world data through \textbf{Fine-tuning}. However, challenges arise in selecting appropriate learning rates, freezing or unfreezing layers, and avoiding overfitting. For example, a developer asked ``\textit{How to train BART for text summarization using custom dataset?}''~\cite{Murugesh2020apr}. Sometimes, developers want to fine-tune PTMs to new tasks. This is more challenging, as these new tasks are often cross-field, not just new datasets in the same field. Thus, major modifications usually need to be made to the model to meet the demand. For example, ``\textit{I fine-tuned a BERT (or RoBERTa) model for sequence classification. Can I fine-tune the same model for a different task (QA or Sentiment Analysis)?}''~\cite{Nilou2021How}.

Another significant challenge that developers face during model training is \textbf{Non-Convergence}. Non-convergence refers to the situation where the model fails to reach an optimal solution during training. This can manifest in various ways, such as the loss function not decreasing, the model's performance plateauing, or the training process getting stuck. For instance, a developer encountered non-convergence when applying a pre-trained HuggingFace ALBERT transformer model to a text classification task, as the loss was not decreasing~\cite{beginner2020ALBERT}.

Compared to DL, training PTMs can be more computationally intensive and time-consuming, making \textbf{Training Acceleration} crucial for efficient model development. Developers often struggle with ways to accelerate the training process. For example, a developer asked, ``\textit{Is there any way to do some CPU optimization to reduce the training time?}''~\cite{k.avinash2020Hugging}. 
Developers also face challenges in \textbf{Understanding Model Training Information}, including concepts and methods. This difficulty highlights the need for clarity and comprehension of training-related information. Additionally, challenges of \textbf{Loss Computation} also arise within this subcategory.

\textit{\textbf{4) Model Loading (4.93\%).}}
The challenge of model loading refers to the difficulties developers may face when loading PTMs into memory. It can be challenging due to factors such as large model sizes, disk I/O bottlenecks, and path issues~\cite{user9010102018Tensorflow}. Developers can address this challenge by considering techniques such as model compression, incremental loading, and model caching. These strategies help optimize the loading process and enhance the integration of PTMs into applications.

\textit{\textbf{5) Memory Management (3.70\%).}}
Developers need to manage memory when working with PTMs to ensure efficient loading, execution, and training. This is crucial due to the large sizes of PTMs, memory requirements during inference and fine-tuning, as well as the demands of batch processing and parallel processing. However, developers often encounter challenges related to memory management, including \textbf{CPU, GPU, or TPU management}. These challenges arise from various factors such as insufficient memory capacity, memory allocation, model parallelism, batch processing, and TPU memory utilization. For instance, a developer asked ``\textit{How can I make full utilization of RAM \& GPU memory?}''~\cite{kalsi2023Low}. Developers can optimize memory management by exploring techniques like data parallelism, memory optimization algorithms, and efficient batch processing. They can leverage memory-efficient tools, frameworks, and hardware-specific optimizations. Effective memory management helps overcome challenges and maximizes resource utilization for PTMs.

\textit{\textbf{6) Other Challenges related to Model Lifecycle Management (6.42\%).}} Besides the challenges discussed earlier, developers may also encounter issues related to \textbf{Model Saving}. Saving a trained model is crucial for future use, deployment, or sharing with others. However, challenges may arise when saving the model weights with customized requirements. These requirements can include saving and averaging weights of models~\cite{Faseela2020TFModelSaving}, or saving only the weights that achieve the best validation performance~\cite{Seewoo2020Save}.
\textbf{Model Testing} is essential to ensure high performance, involving metrics selection, dataset preparation, and comprehensive analysis. Developers may feel puzzled about Model Testing. A developer asked, ``\textit{how to do a test before fine-tuning, in order to compare two models before and after fine-tuning}''~\cite{user152830252021How}. Other challenges include \textbf{Model Deployment} and \textbf{Model Conversion}. Model deployment integrates trained models into a production environment, while Model Conversion converts models between libraries/formats. For instance, developers aim to improve Hugging Face transformer model inference speed without GPU~\cite{DevPy2021Improve} or inquire about converting a BERT model to TFLite~\cite{Ali2020Convert}.

\vspace{1mm}
\noindent \textbf{\textit{II. Coding-related Issues (32.35\%)}}
\vspace{1mm}

When developers utilize PTMs, they may come across various coding-related challenges. These challenges encompass a range of issues, such as correctly using the model's API, ensuring compatibility with the coding environment, handling syntax errors, addressing resource-intensive processing demands, and appropriately addressing warnings. Developers can effectively navigate the complexities of using PTMs by being mindful of potential issues and optimizing their code.

\textit{\textbf{1) Usage of Model API (13.83\%).}}
Developers often encounter challenges related to understanding, generating, or modifying code implementation with PTM APIs. This is a widespread issue that impacts over 10\% of PTM-related questions.
One key difficulty lies in the complexity of PTM APIs, which necessitates a thorough comprehension of their documentation and effective utilization within code. Interacting with a PTM API involves understanding its functionalities, parameters, and input-output data formats. It is essential to have a solid grasp of the underlying structure and functionality of the PTM to effectively work with the API. We find that most of the questions are related to parameters.
For instance, a developer who followed a basic tutorial on the ChatGPT API encountered the following issue: ``\textit{OpenAI ChatGPT (GPT-3.5) API error: `InvalidRequestError: Unrecognized request argument supplied: messages' }''~\cite{corvusMidnight2023OpenAI}. This error indicates that the API request included an unrecognized argument, possibly related to the ``messages'' parameter. Another common issue stems from developers being unclear about the type of parameter expected by an API. For example, a developer encountered the error ``\textit{TypeError: not a string | parameters in AutoTokenizer.from\_pretrained()'' while trying to use the AutoTokenizer.from\_pretrained() function}~\cite{DanielBell992022TypeError}. Additionally, developers may struggle to understand how to use a specific API to implement a particular task. For instance, a developer expressed uncertainty about which API to use when attempting to implement the OpenAI GPT-3 API Client in PHP~\cite{Overstack2022How}. The above challenges indicate that the current API documentation may sometimes be insufficient.

\textit{\textbf{2) Enviornment (11.36\%).}}
Developers may face environment-related issues when working with PTMs, such as compatibility problems with different versions of programming languages, frameworks, or libraries. These issues can arise due to differences in dependencies, package versions, or even operating systems.

\textbf{Library Compatibility/Version.} When working with PTMs, many questions about the Environment often stem from compatibility or version issues with libraries. Incompatibilities between the frameworks or libraries and the PTMs' requirements can cause various problems. Here are some common scenarios. PTMs typically need specific libraries or frameworks to be installed, and using an incompatible version can lead to errors or unexpected behavior. For example, a developer encountered an error (\textit{openai error: subprocess-exited-with-error}) while attempting to install openai with Python 3.11 on Windows OS~\cite{Ljyeon2023How}. The root cause in this case could be the need for a newer version of pip to ensure the proper installation of openai.
Upgrading a library version may also introduce changes to the corresponding API. If developers are not promptly informed of these changes, errors may occur. For instance, when a developer tried to fine-tune a Huggingface GPT2LMHeadModel model for casual language modeling using PyTorch Lightning, they encountered the error: \textit{AttributeError: `str' object has no attribute `size'}~\cite{Nyxynyx2021Huggingface}. The solution to this issue would involve using a recent version of the ``transformer" library, as it likely introduced changes to the API that were not accounted for in the code.

Developers may also face the challenges with \textbf{Library Import}. It means that they do not know how to properly import a certain library. The possible reasons include unfamiliarity with library names, incorrect import syntax, incorrect installation methods, or importing outdated libraries. For instance, a developer may face an issue if they fail to install the ``transformer" library using its official Git link. This can result in errors when attempting to use certain functions from the library~\cite{Laz2020Transformer}. In addition to the challenges mentioned above, developers may also face environmental issues in two other areas. \textbf{Configuration}: Configuration-related challenges involve difficulties in setting up or configuring the PTM for specific tasks or use cases. Developers may need to refer to the PTM's documentation or seek guidance from the community to overcome configuration hurdles. \textbf{Device Issue}: When working with PTMs, device-related challenges can arise due to hardware limitations or compatibility issues. These challenges may include memory constraints, GPU availability, GPU driver version, and optimizing the code for efficient usage of resources. To address these issues, developers may need to adjust batch sizes, utilize distributed computing, or explore hardware-specific optimizations.

\textit{\textbf{3) Other Challenges related to Coding-related Issues (7.16\%).}}
When using PTMs, developers face coding-related challenges beyond the Environment and Usage of Model API. These challenges include \textbf{Syntax Issue}, \textbf{Multi-processing Issue}, and \textbf{Warning Ignoring}. 
Syntax issues result from errors in code structure or format. These issues are often caused by a lack of programming experience or carelessness. Multi-processing challenges arise when developers need to manage the concurrent or parallel execution of code involving PTMs. This can occur in situations where multiple requests must be processed simultaneously or when utilizing distributed computing for quicker inference. For example, a developer used Hugging Face's zero-shot classification pipeline for question answering but found that implementing multiprocessing did not improve the timing compared to using a for loop~\cite{Code_72021Multiprocessing}. Moreover, PTM-related libraries often provide warning messages to alert developers about potential issues or best practices. Sometimes, developers want to ignore these warnings.

\vspace{1mm}
\noindent \textbf{\textit{III. Others (10.86\%)}}
\vspace{1mm}

The third category of PTM-related questions encompasses three subcategories of issues and challenges that typically fall outside the scope of the specific challenges mentioned earlier. 

We observe that 5.68\% of developers propose \textbf{New Requirement} for PTMs. This challenge occurs when developers have specific use cases or requirements that are not currently supported by existing PTM-related APIs. In such cases, developers are interested in understanding how to implement these desired functionalities or features using PTMs. For instance, developers may seek to customize positional embeddings~\cite{Exploring2018How} or integrate PTMs with specific data formats like JSON~\cite{Bob2022OpenAI}, encountering challenges along the way. Another subcategory of challenges, accounting for 4.20\%, is \textbf{Tutorial Seeking}. Developers often face situations where they need to complete a specific task but lack the knowledge or guidance on how to do so. In such cases, they seek tutorials or guidance from the community to overcome these challenges and accomplish their objectives. For example, a developer asked, ``\textit{I would like to train an encoder-decoder model as configured below for a translation task. Could someone guide me as to how I can set up a training pipeline for such a model?}''~\cite{Mitesh2020How}. 
Furthermore, a small percentage of developers (1.23\%) encounter challenges related to \textbf{Model Downloading}. These challenges may involve difficulties in downloading or accessing specific PTM models, which can hinder the progress and usage of the models.

\begin{mdframed}[linecolor=gray,roundcorner=12pt,backgroundcolor=gray!15,linewidth=3pt,innerleftmargin=2pt, leftmargin=0cm,rightmargin=0cm,topline=false,bottomline=false,rightline = false]
  \textbf{Summary for RQ3:} We construct a comprehensive taxonomy of PTM-related challenges, which can be grouped into three categories. The richest category is Model Lifecycle Management, which includes various aspects from model construction to deployment, with nine sub-categories. We also observe that a notable number of developers face coding issues and other challenges such as model downloading. Many challenges are often unique or more serious than those encountered by general deep learning practitioners, such as ``\textit{fine-tuning}'', ``\textit{memory management}'', and ``\textit{model component understanding}''.
\end{mdframed}

\vspace{-0.2cm}
\section{Implications}
\subsection{Implications for Researchers}

We conduct a comprehensive and systematic analysis of the challenges of PTM-related questions on SO, which can help researchers identify current limitations in existing PTMs, as well as inspire new directions and solutions for future research. For example, 
our analysis reveals that many practitioners struggle with understanding the internal workings of PTMs, such as embeddings, weights, and outputs. This suggests that \textbf{more interpretable and explainable PTMs} should be explored, as they can offer greater transparency and insights for users. Additionally, our analysis shows that many practitioners face difficulties customizing or adapting PTMs for new tasks or scenarios, such as changing input or output data formats, adding new training data, or fine-tuning model parameters. This highlights the need for \textbf{more flexible and adaptable PTMs} that can accommodate a wider range of applications and use cases. We also found that incompatibilities between frameworks, libraries, and PTM requirements can pose challenges and hinder the smooth integration and utilization of PTMs. To address this issue, future research could focus on developing \textbf{automated tools to detect and mitigate library conflicts}, as well as providing mechanisms to resolve conflicts and suggest alternative libraries. Moreover, although SO is a representative QA platform for PTM practitioners, future research can  \textbf{supplement our analysis with data from additional platforms and repositories} to obtain a more comprehensive understanding of the PTM domain. 

\vspace{-0.2cm}
\subsection{Implications for Practitioners}
Our taxonomy is a helpful guide for developers who are working with PTMs. They can use the taxonomy combined with our annotated questions to anticipate and overcome specific challenges that they may face during the development and deployment of PTMs. For instance, to tackle challenges in \textbf{model construction}, developers can get guidance on model design, pre-training goals, and selecting suitable PTMs for particular tasks. They can also learn techniques for \textbf{model customization and generalization} to new tasks. In data processing, developers can refer to resources for modifying input data, understanding input data formats, and expanding input data. They can also learn techniques for customizing the model's output and handling output data for downstream tasks. In case of challenges in \textbf{model training}, developers can pay close attention to non-convergence problems and training acceleration on the basis of understanding model training information. They can also learn techniques for managing model loading and optimizing memory usage. In addition to the above, developers can share their experiences, solutions, and insights with the community by \textbf{participating in online forums and engaging with PTM vendors}. They can contribute to the collective knowledge and help others overcome similar challenges. Additionally, our study provides a quantitative analysis of the popularity and difficulty trends of PTM-related questions on SO. This analysis can help practitioners \textbf{stay informed about the latest developments and best practices of PTMs and techniques in the field}.

\vspace{-0.2cm}
\subsection{Implications for PTM Vendors}
The taxonomy of challenges offers valuable insights for PTM vendors to \textbf{improve their offerings and provide better support to developers}. PTM vendors can enhance their products by addressing challenges, such as model loading, memory management, and compatibility with different frameworks and libraries. They can optimize model loading processes, develop memory-efficient algorithms, and provide compatibility guidelines and tools. Additionally, they can \textbf{create comprehensive documentation, tutorials, and resources} to support developers with challenges related to model construction, data processing, and training. By offering clear explanations, code examples, and troubleshooting guides, vendors can help developers work effectively with their PTM offerings. \textbf{Engaging with the developer community} is crucial for PTM vendors. They should establish support channels and bug-reporting mechanisms to listen to developers' needs and challenges. By incorporating feedback and promptly addressing issues, vendors can build strong relationships with developers and foster a collaborative environment for PTM development.

\vspace{-0.2cm}
\subsection{Implications for Educators}
The analysis of challenges related to PTMs on SO has important implications for PTM educators, such as instructors and online course providers. It offers valuable insights into the specific areas where learners may encounter difficulties and can guide the design and delivery of educational materials and courses.
Specifically, PTM educators can \textbf{tailor their curricula} to address the challenges identified in the taxonomy. For example, they can develop instructional materials that focus on explaining the internal workings of PTMs in a more interpretable and explainable manner. This can include providing visualizations, intuitive explanations, and real-world examples to enhance learners' understanding of PTM mechanisms like embeddings, weights, and outputs.
Furthermore, educators can \textbf{design practical exercises and assignments} that allow learners to practice customizing and adapting PTMs for different tasks and scenarios. By incorporating hands-on activities, learners can gain experience in modifying input/output data formats, fine-tuning models, and addressing compatibility issues with libraries and frameworks. Additionally, educators can \textbf{provide resources and guidance on library conflict detection and mitigation}. This can involve teaching learners techniques to identify and resolve conflicts, as well as highlighting best practices for managing dependencies and ensuring compatibility between PTMs and various libraries. By incorporating these insights into their educational materials and courses, PTM educators can better prepare learners to overcome the challenges they may encounter when working with PTMs. This will contribute to a more effective and efficient learning experience, enabling learners to confidently apply PTMs in their own projects and research.

\section{Related Work}

\subsection{Studies on SO Data}

Many studies use SO data to analyze different aspects of software development, such as programming languages \cite{
allamanis2013and}, software testing \cite{
kochhar2016mining}, software security \cite{fischer2017stack}, and software privacy \cite{acar2016you}. Some studies also focus on analyzing specific types of software, such as DL software \cite{
chen2021empirical} and mobile applications~\cite{rosen2016mobile, fontao2018supporting}. Some studies investigate the quality of questions and answers on SO to help developers ask high-quality questions and provide good answers~\cite{ponzanelli2014improving}. Some studies also treat SO as a social platform to explore the social activities of developers~\cite{ blanco2020understanding}.
These studies provide valuable references for the design of our study.

\subsection{Studies on PTMs}
Many studies aim to understand, improve, and apply PTMs. Some studies focus on understanding the behavior and reasoning of PTMs, such as probing their linguistic knowledge \cite{tenney2019bert}, analyzing their attention patterns \cite{ballout2023opening}, and visualizing their hidden states \cite{ballout2023opening}. Some studies have focused on improving the performance and efficiency of PTMs, such as designing better architectures \cite{lan2019albert}, optimizing better objectives \cite{liu1907roberta}, and reducing computation and memory costs \cite{sanh2019distilbert}. Some studies have focused on applying PTMs to various downstream tasks, such as natural language understanding \cite{wang2018glue}, natural language generation \cite{keskar2019ctrl}, and question answering \cite{yoon2019pre}.

Our work complements these studies by providing a practitioner's perspective on the challenges of using PTMs. We collect and analyze real-world questions posted by developers who use PTMs for various purposes. We identify the common problems and difficulties that they encounter when using PTMs, such as selecting the appropriate PTMs, fine-tuning it for specific tasks, deploying it to different platforms, etc. 

\section{Threats to Validity}
\textbf{Selection of data source.} 
Our study refers to the previous work~\cite{chen2020comprehensive, abdellatif2020challenges, haque2020challenges,wang2023automl} to use SO as our data source to extract challenges faced by practitioners in a certain field. It should be acknowledged that it can not cover all the questions posed by practitioners in the field of PTMs. However, as the most popular and representative QA platform in the programming community, leveraging SO's extensive knowledge allows us to gain valuable insights into the PTM domain and contribute to existing research. Our findings can serve as a foundation for future work.

\textbf{Identification of PTM-related questions.}
To include PTM-related questions, we created a set of PTM tags. However, this tag set may introduce bias as it may not cover all PTM-related questions. To address this limitation, we used a two-step approach. Firstly, we followed the methodology of previous work~\cite{chen2020comprehensive} to create an initial set of representative tags. Secondly, we used a snowball sampling technique, leveraging SO's relevant tag recommendation mechanism, to expand our tag set. This iterative process helped reduce the risk of overlooking important PTM-related questions and ensure a more comprehensive coverage of the topic. However, it is worth noting that this approach does not guarantee that all identified questions are PTM-related. As shown in RQ3, around 5.81\% of the questions were unrelated to PTMs. This potential threat may impact the results of RQ1\&2.

\textbf{Sampling of questions for manual analysis.} To select questions for manual analysis in RQ3, we combine random sampling and high-score sampling strategies. While high-score sampling aimed to include questions of high quality and significance, the threshold ($score \geq 9$) for determining high scores is subjective to some extent. This trade-off between the representativeness of challenges and the feasibility of manual analysis is inevitable. It is important to note that the inclusion of high-score questions may slightly deviate from the real-world distribution of challenges.

\textbf{Subjectivity of inspection.} The results of RQ3 may be threatened by the manual analysis process. As with any qualitative research, there is a degree of subjectivity involved in interpreting and categorizing the data~\cite{ratner2002subjectivity}. To mitigate this threat, we employed a rigorous and systematic approach. We established a codebook and clear guidelines for categorization, and the second and third authors were involved to ensure inter-rater reliability. We conducted regular discussions to address any disagreements or uncertainties.

\vspace{-0.1cm}
\section{Conclusion}

This paper presents a comprehensive and systematic analysis of the challenges associated with PTM-related questions on SO. Our findings reveal that PTM-related questions are gaining increasing popularity and are difficult to solve. To gain deeper insights into these challenges, we construct a hierarchical taxonomy consisting of 42 codes and three categories, which provides a structured framework for understanding and addressing the specific challenges encountered in PTMs. Based on our findings, we discuss the implications of our study for various stakeholders, including researchers, practitioners, vendors, and educators in the field of PTMs. To facilitate replication and further research, we make the data available online at \url{https://figshare.com/s/ef756cf1384e4542ddcb}.

\section{Acknowledgements}
This work is supported by the National Natural Science Foundation of China Grants 62202022, 62302021, 62141209, and 62332001, and Self-determined Research Funds of State Key Laboratory of Complex \& Critical Software Environment SKLSDE-2023ZX-15.

\balance
\bibliographystyle{IEEEtran}
\bibliography{ref}

\end{document}